\documentclass[sigconf]{acmart}

\usepackage{booktabs} % For formal tables

\usepackage{siunitx}
\usepackage{multirow}
\usepackage{adjustbox}
\usepackage{color}
\usepackage[utf8]{inputenc}
\usepackage[T1]{fontenc}
\usepackage{xspace}
\usepackage{tabularx}
\usepackage{graphicx}
\usepackage{subcaption}
\usepackage{algorithm}
\usepackage{algorithmic}

\usepackage{enumitem}
\usepackage{balance}

\makeatletter
\DeclareRobustCommand\onedot{\futurelet\@let@token\@onedot}
\def\@onedot{\ifx\@let@token.\else.\null\fi\xspace}

\makeatother

\setcopyright{acmlicensed}
\theoremstyle{definition}

\acmDOI{10.475/123_4}

\acmISBN{123-4567-24-567/08/06}

\copyrightyear{2021}
\acmYear{2021}
\setcopyright{acmlicensed}\acmConference[MM '21]{Proceedings of the 29th
ACM International Conference on Multimedia}{October 20--24, 2021}{Virtual
Event, China}
\acmBooktitle{Proceedings of the 29th ACM International Conference on
Multimedia (MM '21), October 20--24, 2021, Virtual Event, China}
\acmPrice{15.00}
\acmDOI{10.1145/3474085.3475476}
\acmISBN{978-1-4503-8651-7/21/10}

\begin{document}
\title{Assisting News Media Editors with Cohesive Visual Storylines~\textsuperscript{*}}

\thanks{* Please cite the ACM MM 2021 version of this paper.}

\author{Gonçalo Marcelino}
\affiliation{%
  \institution{NOVALINCS}
  \streetaddress{}
  \city{Uni. NOVA de Lisboa, Portugal} 
}
\email{goncalo.bfm@gmail.com}

\author{David Semedo}
\affiliation{%
  \institution{NOVALINCS}
  \streetaddress{}
  \city{Uni. NOVA de Lisboa, Portugal} 
}
\email{df.semedo@fct.unl.pt}

\author{Andr\'e Mour\~ao}
\affiliation{%
  \institution{NOVALINCS}
  \streetaddress{}
  \city{Uni. NOVA de Lisboa, Portugal} }
\email{a.mourao@campus.fct.unl.pt}

\author{Saverio Blasi}
\affiliation{
  \institution{BBC Research and Development}
  \city{London, UK} }
\email{saverio.blasi@bbc.co.uk}

\author{Marta Mrak}
\affiliation{%
  \institution{BBC Research and Development}
  \city{London, UK} }
\email{marta.mrak@bbc.co.uk}

\author{João Magalhães}
\affiliation{%
  \institution{NOVALINCS}
  \city{Uni. NOVA de Lisboa, Portugal} }
\email{jmag@fct.unl.pt}

\renewcommand{\shortauthors}{G. Marcelino et al.}

\begin{abstract}
Creating a cohesive, high-quality, relevant, media story is a challenge that news media editors face on a daily basis.
This challenge is aggravated by the flood of highly-relevant information that is constantly pouring onto the newsroom.
To assist news media editors in this daunting task, this paper proposes a framework to organize news content into cohesive, high-quality, relevant visual storylines.
First, we formalize, in a nonsubjective manner, the concept of visual story transition. 
Leveraging it, we propose four graph based methods of storyline creation, aiming for global story cohesiveness. These where created and implemented to take full advantage of existing graph algorithms, ensuring their correctness and good computational performance. They leverage a strong ensemble-based estimator which was trained to predict story transition quality based on both the semantic and visual features present in the pair of images under scrutiny.
A user study covered a total of 28 curated stories about sports and cultural events. Experiments showed that (i) visual transitions in storylines can be learned with a quality above 90\%, and (ii) the proposed graph methods can produce cohesive storylines with a quality in the range of 88\% to 96\%.
\end{abstract}




\maketitle

\section{Introduction}
Media editors are constantly judging the quality of news material to decide if it deserves being published. The task is highly skilful and deriving a methodology from such process is not straightforward. 
The task of identifying visual material suitable to describe each story segment is, from the perspective of media professionals, highly complex.
The motivation for why some content may be used to illustrate specific segments can derive from a variety of factors. 
Thanks to its widespread adoption, social media services offer a vast amount of available content, both textual and visual, and is therefore ideal to support the creation and illustration of these event stories~\cite{paulussen2008user,Hu:2012:BNT:2207676.2208672,Doggett2016IdentifyingEN,Tolmie2017SupportingTU}. Unfortunately though, exploiting this information is not easy, in that media professionals are faced with the daunting task of needing to manually search over a massive amount of data~\cite{diakopoulos2012finding}, which can be resource expensive, time consuming and lead to unsatisfactory results. Finding ways to support or automate this process would, therefore, be highly beneficial. This is not an easy task because such systems would need to (1) find appropriate information to illustrate a news timeline using camera crew and newsroom specific social media tools, and (2) make sense (from an editorial perspective) of the timeline of events across different textual and/or visual documents.

\begin{figure}[h]
    \centering
    \includegraphics[width=1\linewidth]{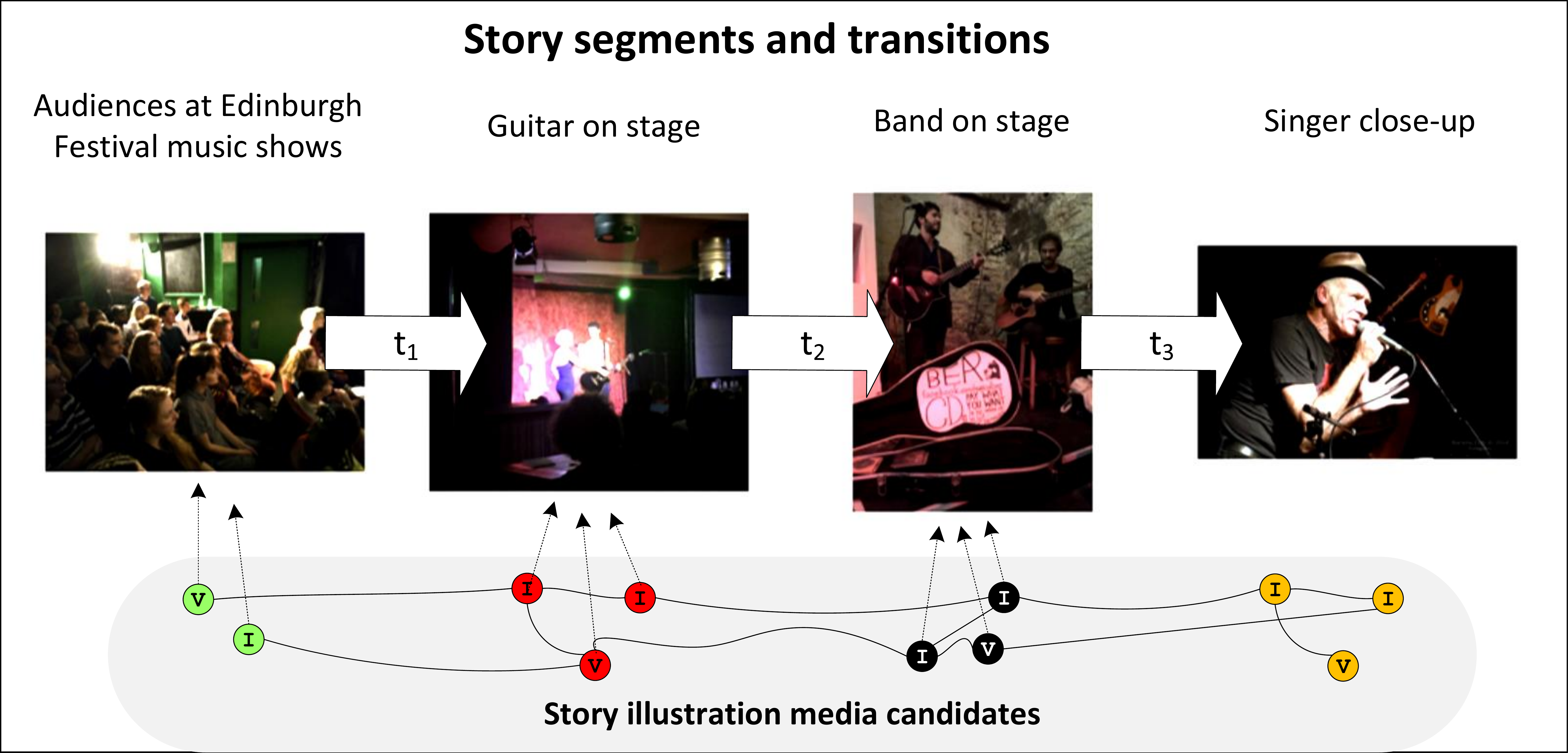}
    \caption{Visual storyline editing task: a news \textit{story topic} and \textit{story segments} can be illustrated by social-media content.}
    \label{fig:social-visual-story}
\end{figure}

These timelines, illustrated in Figure~\ref{fig:social-visual-story}, refer to a story topic and are structured into \textit{story segments} that should describe narrow occurrences over the course of the event. In practice, \textbf{the task of the media editor is to illustrate each story segment by selecting the most cohesive set images/videos from a set of candidate images/videos}. In this setting, we propose a graph-based framework to organize content into a \textit{cohesive visual storyline}. 
The first contribution tackles a non-trivial task: mimicking the human perception of \textit{visual story transition quality}.  
To do so, we propose a novel formal definition of transition between two media documents. Leveraging this definition we are then able to study the impact of a large set of  semantic and visual criteria in the quality of image transitions.
This proposal is utterly important as it allows predicting the perceived transition quality of pairs of images by analyzing their visual and semantic characteristics. 
The second contribution moves beyond individual story transitions and looks at the storyline as a whole. We propose a set of multimodal graph based methods that, given a story with N segments, and a list of sets of candidate multimodal documents to illustrate each segment, produce visual storylines composed of images.

In summary, the proposed framework targets the media professional who needs to find content to illustrate a news story. Our framework works as follow: (1) the media editor submits a story topic and its story segments as a query; (2) the framework computes a sequence of story illustrations in a cohesive way; and (3) this rank of story illustrations is then presented to the media editor for selection and post-editing.
Hence, the gain is obvious: instead of struggling with a wide range of content permutations to illustrate story segments, the media editor can now browse different sequences of story illustrations and work from there.

%-------------------------------------------------------------------------
\section{Related Work}
From our discussions with media professionals, our goal was to design algorithms that could assist them in illustrating event plots. In this context, one of the few works that provide insight into the task is \cite{delgado2010assisted}. In it, the authors observe that presenting repeated images/videos in the context of news related storylines, makes viewers perceive said storylines as having less quality, even if relevant information is being shown.
Strong transitions~\cite{Iyyer_2017_CVPR, Piacenza2011} to cause a laugh, surprise, fear, etc., can be introduced by editors, who always have the final decision, by modifying candidate illustrations. 

%----- Timeline generation
\textbf{Collaborative Timelines.} A highly appealing characteristic of social media platforms is that they make available large amounts of real-time data (news updates, opinions and first-hand reports) about specific events~\cite{McMinn:2013:BLC:2505515.2505695} in a streaming fashion, including important temporal clues about shared content~\cite{10.1145/3343031.3351036,10.1145/3394171.3413540}. Hence, the synchronisation and organisation of collaborative media along the temporal dimension leads to the generation of event timelines. TwitInfo~\cite{Marcus2011} is a system that displays information about an event, sub-divided into several sub-events identified by automatically-detected peaks. Peak detection is also employed by~\citet{Nichols2012} to detect and summarise important updates to an event, and by~\citet{Lin2012} to build summaries in the form of storylines. \citet{Wang2014} propose to generate summaries and timelines for continuous social media streams with an online clustering algorithm, maintaining distilled statistics in a data structure later used for summarization.
~\citet{10.1145/3394171.3413540} learn a temporal cross-modal embedding space, enabling the creation of multimodal summaries by navigating over that space.
\citet{FreddyChongTatChua2013} propose a topic model that explores the correlation among posts published around the same time, extracting topics that identify different episodes of the same event. The event is summarised by displaying the most representative post for each episode.

To enrich summaries with visual information, media needs to be processed and organized based on its content. 
Graph-based approaches \cite{Schinas2015, kim2014reconstructing, yang2012photo, Kim2014} are among the most popular methods for summarising collaborative visual streams.
\citet{yang2012photo} focused on the alignment of photo streams in a common timeline using a bipartite kernel sparse representation graph. A master stream, corresponding to the summary, is obtained by removing redundant photos.
The construction of large-scale storyline graphs is the central challenge addressed by~\citet{kim2014reconstructing}. The optimisation framework infers a time-varying directed storyline graph that supports multiple storyline branches. 
\citet{Kim2014} propose to use videos to model the sequence of actions that are more likely to occur on a given storyline. These sequences are captured on a sparse graph to later help the temporal grounding of images and the reconstruction of the timeline.
\citet{Schinas2015} propose to use a graph-based approach to detect topics across a dataset of microblog messages organised in a multi-graph, followed by ranking the images embedded in those messages according to relevance and diversity, to produce a visual summary.

\textbf{Qualitative attributes in photos and films.} Solving the problem of information relevance is hardly enough, in the creation of visual summaries that matches user expectations. Past work in visual quality~\cite{luo2008photo}, and more specifically in  visual aesthetics~\cite{dhar2011high} and  interestingness~\cite{gygli2013interestingness}, have proven that subjective visual attributes can computationally modelled and be applied to improve user experience. In illustrating visual storylines, visual memorability~\cite{isola2011understanding} plays an important role in the narrative of the story. For example, it has been observed~\cite{delgado2010assisted} that repeating images/videos in a story is not appreciated, even if relevant information is being shown.

In \cite{Nichols2012}, Nichols et. al. propose a methodology to perform automated text summarization of sporting events using data collected from Twitter. In this case, the authors find the most important moments of an event by identifying spikes in the volume of tweets over time, an approach also applied in TwitInfo~\cite{marcus2011twitinfo}. After acquiring these tweets and applying spam removal techniques to the set, a phrase graph is created from the text present in the tweets. This graph details the chance of a word appearing next to a previously established sequence of words in the context of a phrase. The phrase graph is used to generate possible sentences that summarize the event. These sentences are also scored through the phrase graph and the best scoring ones are outputted. As a possible alternative to this graph based algorithm, the authors propose Sharifi's modified TF.IDF \cite{sharifi2010experiments} as a method for generating a text summary of the event. Although not directly related to visual storyline generation, these works highlight the importance of considering the amount of content being posted to social media, at any particular time, as metric for finding interesting and important events. 

Three years after Nichols et. al. published \cite{Nichols2012}, the authors of \cite{Schinas2015}, Schinas et. al., tackled a similar the problem, by rooting their research in a similar approach. In this work the task of visual event summarization using social media content, again from Twitter, is tackled through the use of topic modelling and graph based algorithms. The authors start by filtering a stream of tweets from a specific event in order to obtain only the most informative ones. This filtering process takes into account image size, image type, text size, text morphology through part-of-speech tagging, among other criteria, leading to the creation of a multigraph. 
Re-posts are then discarded and tweets with duplicated images are clustered with the help of the graph. The tweets in these clusters are removed from the graph and are replaced by a single node that encompasses the information of the removed tweets. The authors then intent on discovering which tweets are part of the targeted event. In order to do this, they apply the SCAN \cite{xu2007scan} algorithm to the graph, which returns a set of subgraphs, each corresponding to a different topic. Finally, for each subgraph, the authors extract the respective images and rank them according to their popularity, relevance to the topic and the amount of information they provide.

In \cite{Mcparlane2014} the authors also tackle the problem of visual summarization, although in this case the summarization is done with images not only from Twitter but also from other websites found in the URLs of tweets. After filtering the obtained images in a manner similar to the one proposed in \cite{schinas2015visual}, the authors rank the remaining images using social signals. Finally, besides taking into account the popularity of an image the authors also factor into its rank the diversity of said image when compared to the remaining available ones.
Of note is that both authors of \cite{schinas2015visual} and \cite{Mcparlane2014} tackle the problem of filtering duplicated content, although through different methodologies. In order to reconcile both approaches one could use pHash as proposed by \cite{Mcparlane2014} and cluster the results through a graph based approach in a way similar to what is described in \cite{schinas2015visual}.

\section{Preliminaries}
\label{sec:candidates}
A visual storyline must be succinct and cohesive but also pleasing to the viewer. Overall it must present a sequence of images/videos as an interesting and relevant narrative. 
Before building the story graph, we first need to retrieve content that is relevant to the different story segments.
In practice, a media editor defines a story composed of $N$ segments, $Story_N=(u_1, u_2, ..., u_N)$, that need to be illustrated ($u_i$). The segments $(u_1, u_2, ..., u_N)$ are then used to query the set of media documents, $D$, and the BM25 retrieval method produces a list of sets of candidate images/videos $(C_1, C_2, ..., C_N)$, where $C_i=\{a_1, ..., a_k\}$ contains candidate images $a_j$ to illustrate the story segment $u_i$.

\section{Multimodal Story Graphs}
\label{sec:storyline_generation}
We now tackle the task of automatic visual storyline illustration. 
This method takes as input, sets of candidate images $C_i$ to illustrate each segment $u_i$ of a story and is tasked with outputting visually and semantically cohesive storylines composed of images in the said sets. We propose two different multimodal graph based approaches to tackle this problem, each with two variants.

\subsection{Scoring Multimodal Story Transitions}
\label{sec:transtions}
Rating a transition between a pair of images according to its quality is a non-linear process that results from the interpretation of the features of the individual images and of the manner in which they interact. 
To tackle the task of optimizing the transition quality between two pairs of images we first need a computationally valid approach to describe the concept of transition. From a non-computational, professional perspective, literature characterizes transitions based on the relations between semantic and visual characteristics of the images that compose them and by the ways in which these images interact. We emulate this approach proposing a novel formalization of transition based on the concept of a distance ensemble. 

To tackle the automation of this process we resort to Gradient Boosted Regression Trees~\cite{mohan2011web, friedman2001greedy}, and define the problem as one of predicting a rating given the set of \textit{transition distances} between a given pair.
Aiming to build a robust model, we propose the use of large set of features to compose the transitions' distances between image pairs. 
More specifically, we define the \textit{story transition quality estimator} as the Gradient Boosted Regressor (GBR) 
\begin{equation}
    transQuality(a,b) \in [0,1],
\label{eq:trans_quality}
\end{equation}
for a pair of any two images $a$ and $b$. 
In practice the GBR needs to estimate a transition quality score between images $a$ and $b$, from a set of \textit{transition distances} $ D=\{\forall c \in I_f, dist_c(feat_c(a), feat_c(b))\}$,
where $I_f$ is the set of features under consideration.
Hence, a transition between two images is formalized as an ensemble of regression trees over a set of distances between the features of said images. 
These features $I_f$ are presented in Table~\ref{tab:transition-features}. 
As training data we considered the transition quality ground truth provided by human annotators.

\begin{table}
\centering
\renewcommand{\arraystretch}{1.8}
\begin{adjustbox}{width=1\columnwidth}
\begin{tabular}{llp{5cm}}
\toprule
Feature Name ($p$) & $distance_p(f_1, f_2)$ & $feature_p(a)$ \\
\midrule
Luminance & $ abs(f_1 - f_2) $ & A positive real value representing the luminance.\\
Color histogram & $\sum abs(f_1 - f_2)$ & A 3D color histogram with 16 bins per RGB channel converted to CIELAB color space.\\
Color moment & $euclidean(f_1, f_2)$ & A vector representing the first color moment of the image in CIELAB color space.\\
Color correlogram & $\sum abs(f_1 - f_2)$ & A 16 bins 3D color correlogram in CIELAB color space.\\
Entropy & $ abs(f_1 - f_2) $ & A positve real value representing the entropy of the image.\\
\#Edges & $ \sum abs(f_1 - f_2) $ & A vector containing the number of horizontal, vertical and diagonal edges.\\
pHash & $ hamming(f_1, f_2) $ & A pHash vector.\\
\midrule
Concepts & $ \#(f_1 \cap f_2) $ & A set of image concepts extracted using VGG16.\\
CNN Dense & $euclidean(f_1, f_2)$ & The embeddings extracted from the last layer of the ResNet CNN.\\
Environment & $f_1 = f_2$ & Either "outdoors" or "indoors".\\
Scene category & $\#(f_1 \cap f_2)$ & The location depicted in an image described through labels (e.g.: "bridge", "forest path", "skyscraper", etc.).\\
Scene attributes & $\#(f_1 \cap f_2)$ & The attributes of the location depicted in an image described through labels (e.g.: "man-made", "open area", "natural light", etc.).\\
\bottomrule
\end{tabular}
\end{adjustbox}
\caption{The multimodal set of features $I_f$, corresponding distance functions and descriptions.}
\label{tab:transition-features}
\vspace{-5mm}
\end{table}

\subsection{MaxTransitions: Bi-partite graph}
This approach optimizes for storylines with the best possible sequential transitions. In this context, the transition quality is only measured between the candidate images of consecutive segments. 

We follow a graph based approach to tackle this problem.  
Specifically, we define $G=(V, E)$, as a \textit{sequence of bipartite weighted directed graphs}, where given a story $Story_N=(u_1, u_2, ..., u_N)$ of $N$ segments, the graph $G$ is constructed as follows:
\begin{enumerate}
\item \textbf{Vertices:} the graph's vertices $V$ correspond to all the candidate images in the sets $(C_1, C_2, ..., C_N)$, of $k$ images each. Each set $C_i=\{a_1, ..., a_k\}$, corresponds to a story segment $u_i$. Hence, each candidate image $a_*$ of candidate set $C_i$ becomes a vertex $v_{i}$ in the graph;

\item \textbf{Edges:} the graph's edges $E$, associate all the candidate images from neighbouring candidate sets. In other words, all vertices in set $C_i$ are fully connected and directed to vertices in set $C_{i+1}$. Hence, the bipartite property of graph;

\item \textbf{Edges-weight:} the weight associated with each edge $e \in E$, connecting two vertices $v_1$ and $v_2$ is given by a function $pairCost(v_1, v_2)$ 
\end{enumerate}

Leveraging this graph structure, exemplified in Figure~\ref{fig:second_sequencial_graph}, for a $N=4$ segment story, we propose two different story transition methods to extract the visual storyline.

\subsubsection{Shortest-Path Solution}
The first approach, regarded as \textit{Shortest Path without relevance} ($MaxTransition$), is optimized for creating storylines with the best possible sequential transitions, regardless of the relevance of the candidate images to the segments they attempt to illustrate. It is designed to present added value in situations where most or all candidate images are already highly relevant to their respective segments or where content relevance is not the highest priority in the context of the story being illustrated.

Formally, this first proposed method computes the shortest path of size $N$ in the graph (i.e. the path will contain a number of vertices equal to the number of segments in the story being illustrated). Hence, we aim to minimize the following expression:
\begin{equation}
\label{eq:sect}
    \min_{v_1 \in C_1, v_2 \in C_2, \dots, v_N \in C_N} \sum_{i=1}^{N-1} pairCost(v_i, v_{i+1})
\end{equation}
where $pairCost(v_x, v_y) = transQ(v_x, v_y)$ and \begin{equation}
transQ(v_x, v_y) = 1 - transQuality(v_x, v_y)
\end{equation}
the function $transQuality$ in equation~\ref{eq:trans_quality}.

The resulting storyline from this approach is the one composed by the images represented by the vertices that minimize expression \ref{eq:sect}. In practice we resort to a variation of Dijkstra's minimum cost path algorithm to solve this problem.

\subsubsection{Weighted Shortest-Path}
The previous alternative considers only transition quality when generating visual storylines. For situations where relevant content is scarce, we propose a second approach, \textit{Sequential with relevance} ($MaxTransitionsRel$). In such situations, some of the available candidate images might not be relevant to the segment they attempt to illustrate. Hence, we leverage both transition quality and relevance of the candidate images, encouraging the creation of storylines with the most relevant candidate images that also present quality transitions.

\begin{figure}
\centering
\begin{subfigure}{.48\columnwidth}
    \centering
    \includegraphics[width=\columnwidth]{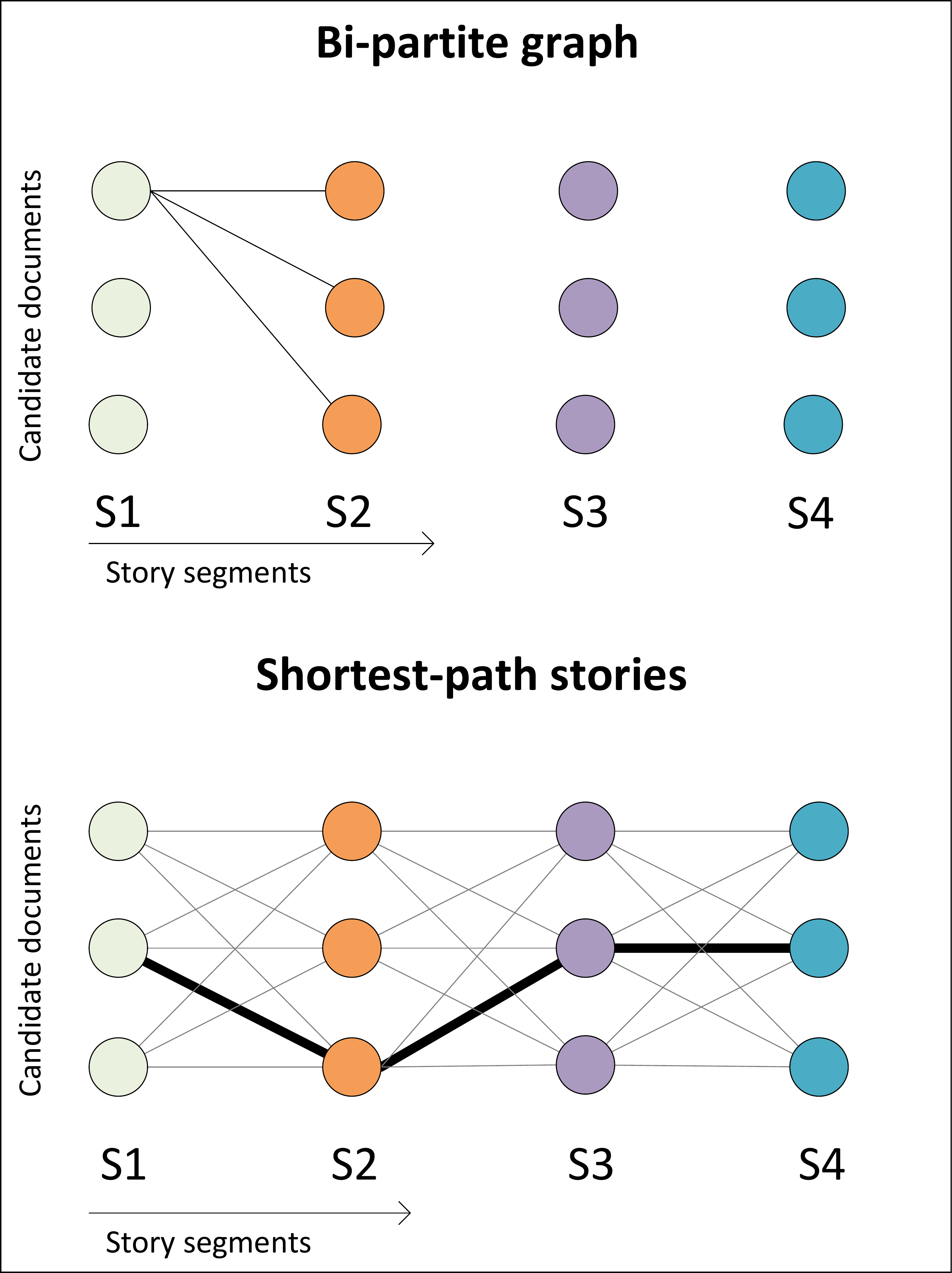}
    \caption{Bi-partite graph.}
    \label{fig:second_sequencial_graph}
\end{subfigure}%
\begin{subfigure}{.492\columnwidth}
    \centering
    \includegraphics[width=\columnwidth]{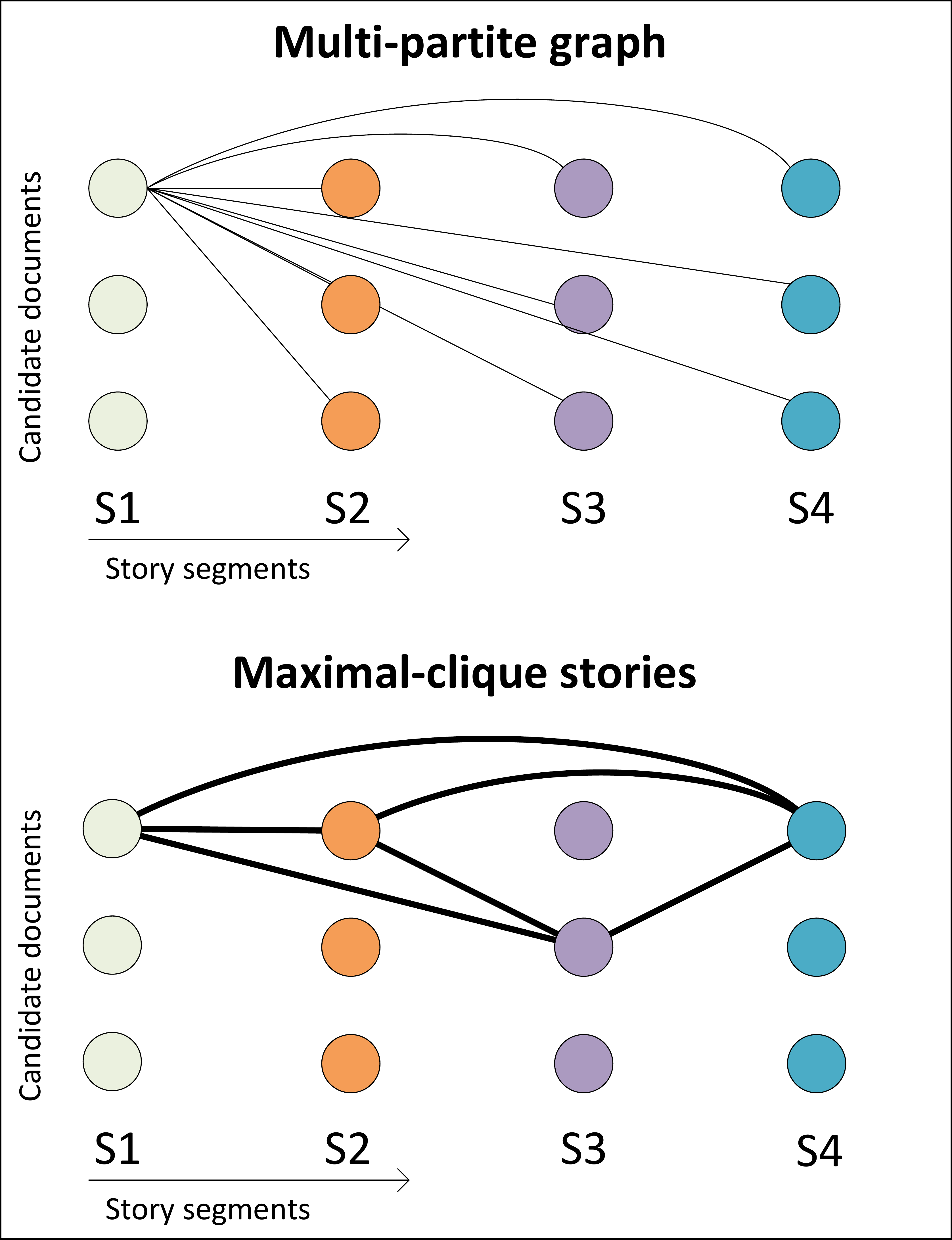}
    \caption{Multipartite graph.}
    \label{fig:fully_connected_graph}
\end{subfigure}%
\caption{Media is represented by the vertices of the graph, each vertex belonging to a candidate set $C_i$. The cost associated with an edge directed from vertices $v_x$ to $v_y$ is given by the $pairCost(v_x, v_y)$ function. The bi-partite graph connects all media in one segment to all media in the \textit{next segment} -- the shortest path finds the solution with maximal similarity between neighboured segments.
The Multipartite graph connects all media in one segment to all media in the \textit{other segments} -- the maximal clique finds the solution with maximal similarity between all documents.}
\label{fig:graph}
\end{figure}

Thus, for this approach, we again aim to find a path composed of $N$ vertices in the graph. However, this time, the expression to minimize weighs both the importance of the first segment in the storyline being relevant, as well as the importance of transitions \textit{vs.} relevance, to score overall storyline quality. We do so by basing ourselves in the transition quality estimator proposed in section~\ref{sec:transtions}. In practice, we attain this path by minimizing the following expression:
\begin{equation}
\begin{split}
\label{eq:sectr}
    \min_{v_1 \in C_1, v_2 \in C_2, \dots, v_N \in C_N} 
    &\beta \cdot \frac{1}{2(N-1)} \cdot \sum_{i=1}^{N-1} pairCost(v_i, v_{i+1}) \\ &+ (1-\beta) \cdot relC(v_1)
\end{split}
\end{equation}
where $\beta$ is used to control the importance of the first segment, and the function $pairCost(v_x, v_y)$, that ranges from 0 to 2, is defined as:
\begin{equation}
\begin{split}
pairCost(v_x, v_y) &= \underbrace{(1-\gamma) \cdot (relC(v_x) + relC(v_y))}_\text{segments illustration} 
\\&+ \underbrace{\gamma \cdot (relC(v_x) \cdot relC(v_y) + transQ(v_x, v_y))}_\text{transition}
\label{eq:pairCostR}
\end{split}
\end{equation}
capturing segment illustration relevance in the first term, and accounting for relevance and quality of the transition, in the second term. $\gamma$ controls this trade-off.
Here, $relC(v) = 1 - rel(c)$. In turn, $rel(c)$ is the normalized relevance of the candidate image represented by vertex $v$ to the text segment $u_i$ to be illustrated, as calculated through the BM25 retrieval method.

The resulting storyline is the one composed by the images represented by the vertices that minimize expression \ref{eq:sectr}.

\subsection{MaxCohesion: Multipartite graph}
The \textit{MaxTransition} approach aims to produce storylines with high transition quality for sequential pairs of images. However, a visual storyline is consumed as a whole by it's viewers, not as a disconnected set of pairs. Consequently we posit a second approach, \textit{Fully connected}, designed to ensure maximal cohesion between all elements of the generated visual storylines, leveraging the possibility that individual transition quality is affected by the remaining elements of the storyline they are part of. Hence, we need to compute the transitions between all elements of the storyline to ensure maximal cohesion.

Again, we follow a graph based approach to tackle this problem.  %Figure~\ref{fig:second_sequencial_graph} depicts the graph
We define $G=(V, E)$, \textit{a N-partite weighted graph}, where given a story $Story_N=(u_1, u_2, ..., u_N)$ of $N$ segments, the graph $G$ is constructed as follow:

\begin{enumerate}
\item \textbf{Vertices:} the graph's vertices $V$ correspond to all the candidate images in the sets $(C_1, C_2, ..., C_N)$, of $k$ images each. Each set $C_i=\{a_1, ..., a_k\}$, corresponds to a story segment $u_i$. Hence, each candidate image $a_*$ of candidate set $C_i$ becomes a vertex $v_{i}$ in the graph;

\item \textbf{Edges:} the graph's edges $E$, associate all the candidate images from \emph{all other candidate sets}. In other words, all vertices in set $C_i$ are connected to vertices in all the candidate sets except $C_i$. Hence, the multipartite property of the graph;

\item \textbf{Edges-weight:} the weight associated with each edge $e \in E$, connecting two vertices $v_1$ and $v_2$ is given by a  function $pairCost(v_1, v_2)$
\end{enumerate}

This graph construct, exemplified in Figure~\ref{fig:fully_connected_graph} for a $N=4$ segment story, is well suited to extract storylines with maximal cohesion as we are going to see next.

\subsubsection{Maximal-Clique Solution}
First, we optimize for maximal cohesion, i.e. for maximal similarity among all story illustrations. Hence, for this first \textit{MaxCohesion} approach, we compute the maximal weighted clique containing $N$ vertices of graph $G$. (i.e. the clique will contain a number of vertices equal to the number of segments in the story being illustrated). Additionally, we pose the following restriction to the clique: it can only contain one vertex per candidate set. 
The bottom graph of Figure~\ref{fig:fully_connected_graph} provides an example of such a clique.

We attain this clique by minimizing the following expression:
\begin{equation}
\label{eq:fult}
    \min_{v_1 \in C_1, v_2 \in C_2, \dots, v_N \in C_N} \sum_{i=1}^{N-1}\sum_{k=i+1}^{N} pairCost(v_i, v_k)
\end{equation}

The resulting storyline from this approach is the one composed by the images represented by the vertices that minimize expression \ref{eq:fult}. We refer to this approach as \textit{maximal clique without relevance} ($MaxCohesion$).

\subsubsection{Weighted Maximum-Clique Solution}
This final alternative $MaxCohesionRel$ builds on the previous one, ensuring high transition quality between all pairs of images in the resulting visual storylines, not just sequential pairs, while optimizing for relevance as with the $MaxTransitionsRel$ approach.

To do so, we compute a weighted clique containing $N$ vertices of graph $G$, again with the restriction that it can only contain one vertex per set of candidate images. To find this clique we minimize the following expression.
\begin{equation}
\label{eq:fultr}
\begin{split}
    \min_{v_1 \in C_1, v_2 \in C_2, \dots, v_N \in C_N} &\beta \cdot \frac{1}{N(N-1)} \cdot \sum_{i=1}^{N-1}\sum_{k=i+1}^{N} pairCost(v_i, v_k) \\ &+ (1-\beta) \cdot relC(v_1) 
\end{split}
\end{equation}
where the function $pairCost(v_x, v_y)$, a function that ranges from 0 to 2, is defined in Eq~\ref{eq:pairCostR}.
The resulting storyline from this approach is the one composed by the images represented by the vertices that minimize equation \ref{eq:fultr}. We refer to this final method as \textit{Fully connected with relevance ($MaxCohesionRel$)}.

\section{Evaluation}

\subsection{Datasets}
Events adequate for storytelling were selected, namely those with strong social-dynamics in terms of temporal variations with respect to their semantics (textual vocabulary and visual content). 
News story topics and their segments were defined collaboratively by media editors and the authors of this paper. These stories were related to events usually covered in the newsroom. News and transitions were evaluated by newsreaders. As annotators, we used 5 newsreaders with ages in the range of 23 to 29 years old~\cite{SocialStories}.

\subsubsection{The Edinburgh Festival (EdFest)} consists of a celebration of the performing arts, gathering dance, opera, music and theatre performers from all over the world. The event takes place in Edinburgh, Scotland and has a duration of 3 weeks in August.

\subsubsection{Le Tour de France (TDF)} is one of the main road cycling race competitions. The event takes place in France (16 days), Spain (1 day), Andorra (3 days) and Switzerland (3 days).

\subsubsection{Data Filtering and Curation}
Having gathered all tweets relevant to the topic and posted within an event's time frame, we proceed to filter documents unsuitable for summarization, according to several quality criteria~\cite{RankingQuality}:

\begin{description}

\item[Retweets.] Since retweets consist of a copy of an original tweet, they are redundant, thus they are excluded.

\item[Language filtering.] The Twitter language filter was already used, i.e., only tweets with text that Twitter deemed as English were collected. A second language filter is applied to exclude Non-English tweets using the LangDetect library~\footnote{\url{https://pypi.python.org/pypi/langdetect}}.

\item[SPAM] In social-media platforms like Twitter, spam is abundant and a serious issue. Thus, we apply a set of filtering methods in an attempt to clean the collected data. We follow the approach in~\cite{McMinn:2013:BLC:2505515.2505695} and excluded tweets containing more than 3 hashtags, more than 3 mentions or more than 2 URLs. We adopt these empirical values which were chosen by the authors due to~\cite{benevenuto2010detecting}.

\item[Visual SPAM.] To detect synthetic and captioned visual elements, we train a logistic regression classifier (with L1 penalty and $\lambda = 1.0$) that predicts if an image is a photograph or not. 
Features include luminance and edge histograms of each image (with edges grouped into vertical, horizontal, 45 and 135 degrees orientations). We employ 5-fold cross-validation and train the model with the dataset from \cite{wang2006npic} plus extra data identified by us. Additionally, we perform OCR to detect and exclude visual elements containing text using the PyTesseract\footnote{\url{https://github.com/madmaze/pytesseract}} library.
See~\cite{RankingQuality} for details.
\end{description}

\subsubsection{Story Transitions ground truth} 
Target storylines and segments were obtained using several methods, resulting in a total of 40 generated storylines (20 for each event), each comprising 3 to 4 segments. In practice, the 320 distinct visual  (232 composed of 4 segments and the remaining 88 composed of 3 segments) were presented to annotators and were asked to rate (i) each segment illustration as relevant or non-relevant, (ii) the transitions between each of the segments in a Likert-scale of 1 to 5, and finally, (iii) the overall story quality, also in Likert scale. Stories were visualized and assessed in a specifically designed prototype interface. It presents media in a sequential manner to create the right \textit{story} mindset to the user.
By taking advantage of the annotations made to the 320 storylines we attained ground truth for a total of 872 pairs of images regarding transition quality, by averaging the scores (0 or 1) provided by the annotators to each pair.

\begin{figure}[t]
    \centering
    \includegraphics[trim={50 0 50 0},clip,width=0.32\columnwidth]{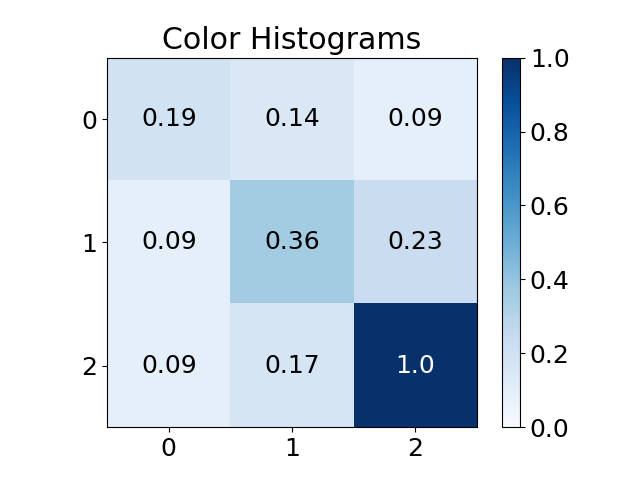}
    \includegraphics[trim={50 0 50 0},clip,width=0.32\columnwidth]{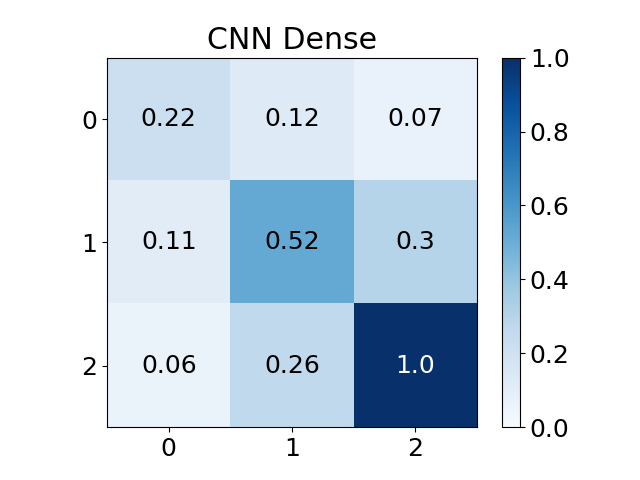}
    \includegraphics[trim={50 0 50 0},clip,width=0.32\columnwidth]{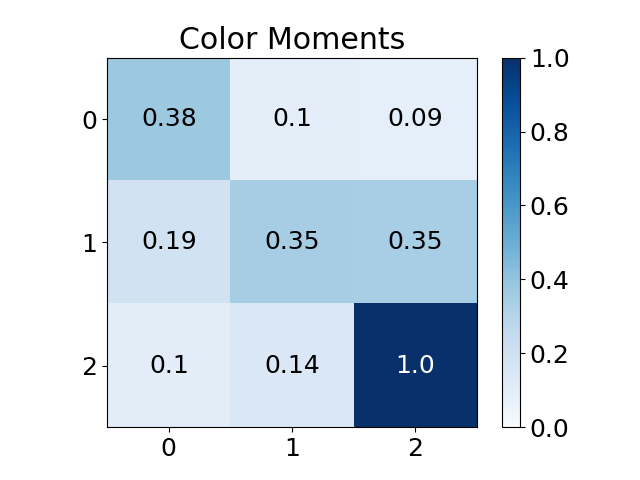}\\
    \includegraphics[trim={50 0 50 0},clip,width=0.32\columnwidth]{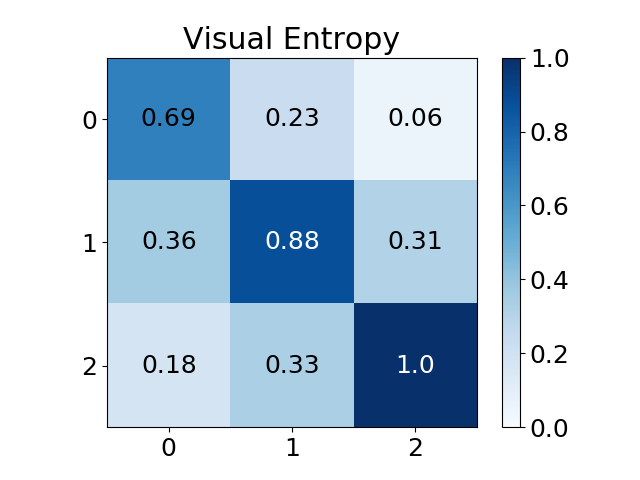}
    \includegraphics[trim={50 0 50 0},clip,width=0.32\columnwidth]{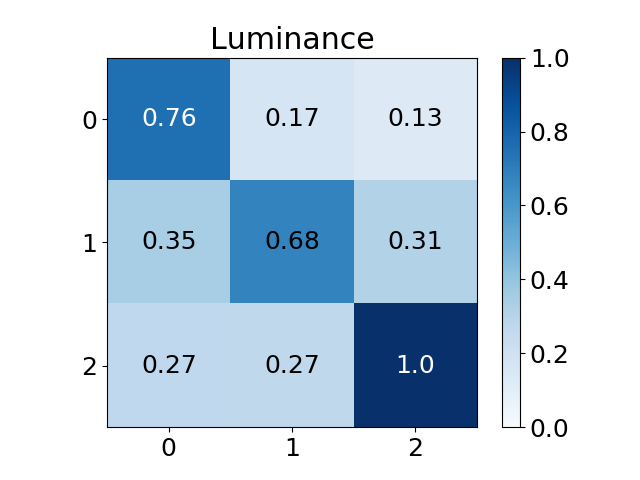}
    \includegraphics[trim={50 0 50 0},clip, width=0.32\columnwidth]{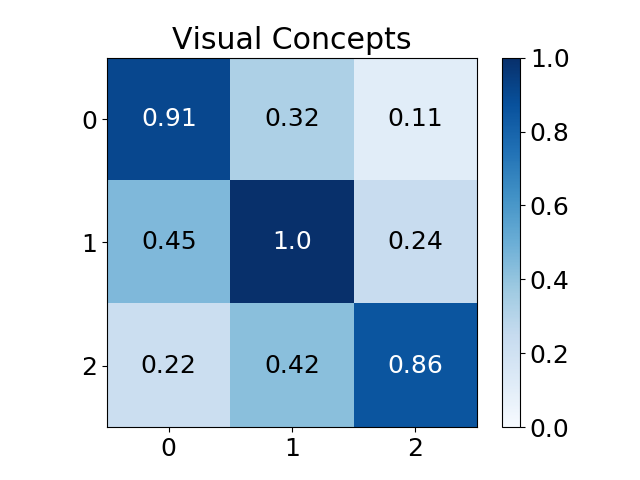}
    \caption{Story transition confusion matrices for different of features. On each matrix, the y-axis regards the score provided by the individual annotators and the x-axis regards the score attained through majority vote. The diagonal is clearly visible for all features, hence, showing a good user agreement.}
    \label{fig:transitions_performance}
    \centering
\end{figure}

\begin{figure}[t]
    \centering
    \includegraphics[width=\linewidth]{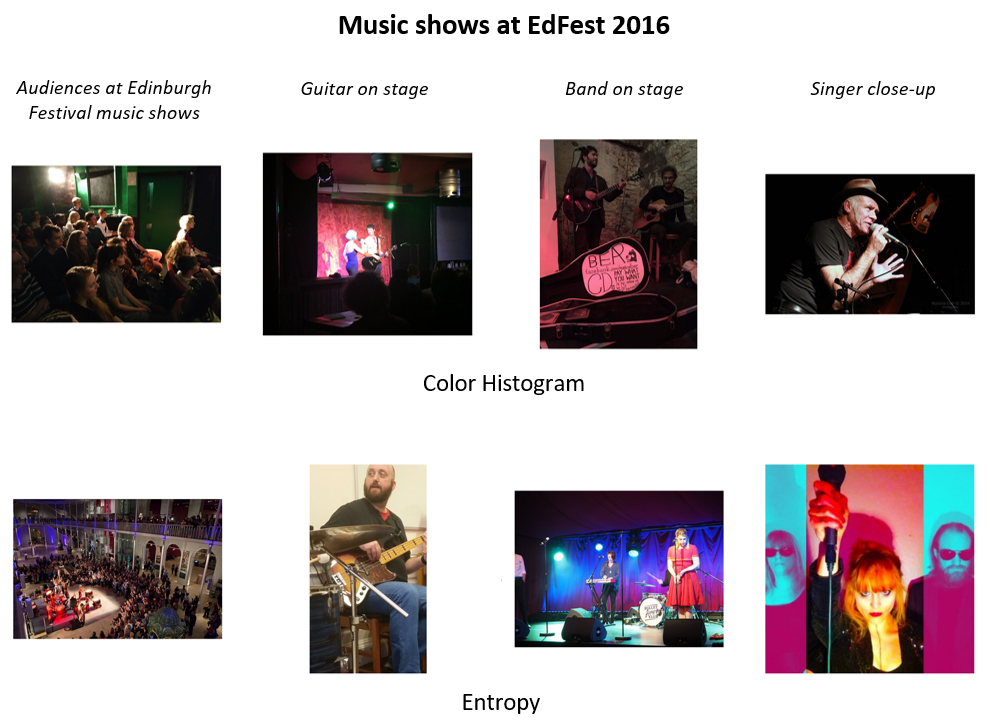}
    \vspace{10mm}
    % \centering
    \includegraphics[width=\linewidth]{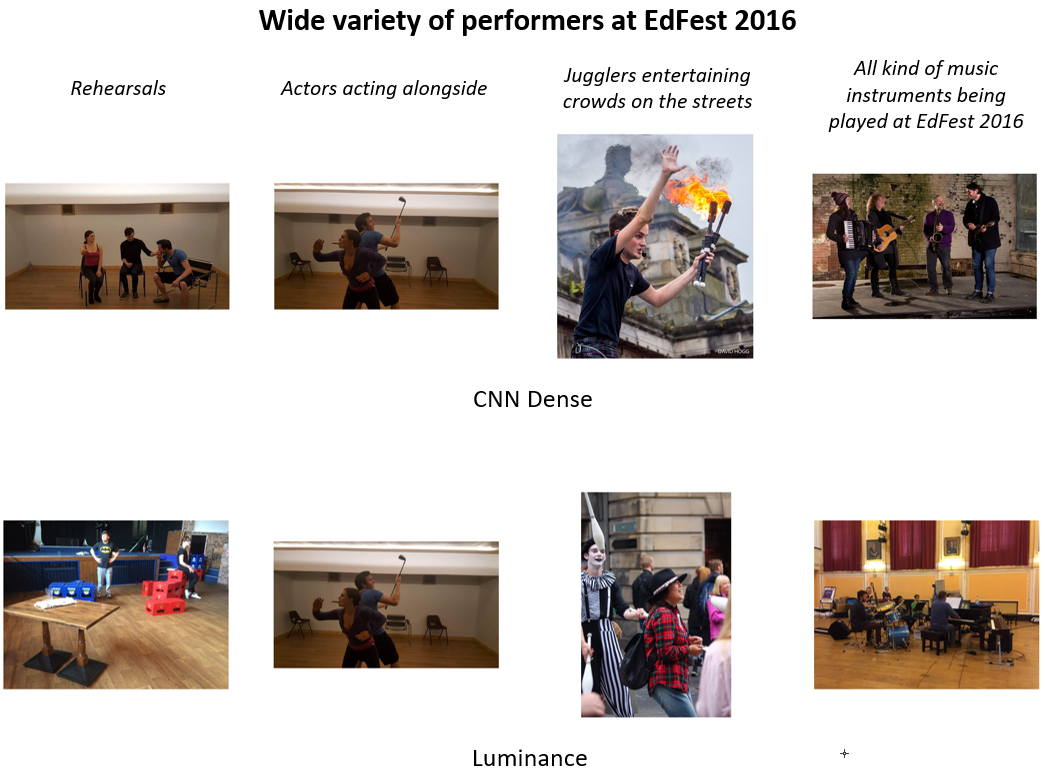}
    \vspace{-10mm}
    \caption{Storyline transitions for two different stories. 
    On the top story example, the transitions of the storyline created with the \textit{Color histogram} baseline obtained an average score of 1 while the ones in the storyline created by the \textit{Entropy} baseline obtained an average score of 0.6. 
    On the bottom story example, the transitions of the storyline created with the \textit{CNN Dense} baseline obtained an average score of 0.93 while the ones in the storyline created by the \textit{Luminance} baseline obtained an average score of 0.87.
    }
    \vspace{-3mm}
    \label{fig:transition-examples2}
\end{figure}

\subsection{User Study: Multimodal Story Transitions}
To understand which individual features have more impact in the story transition quality, we conducted a user study.
The protocol is as follows. To ensure that no irrelevant content would interfere with the assessment of story transitions, we looked at the ground truth and selected a total of 1572 relevant images to illustrate the segments of the 40 stories related to the EdFest and TDF datasets and then sorted them by their BM-25 score. This corresponds to an average of 10 relevant candidates for each story segment. Afterwards, each story was illustrated by considering the ~10 relevant images per segment, and selecting the segment's illustration sequence that minimizes the sum of the pairwise \textit{transition distances}, composed only of a single feature, between sequential images. Each baseline focuses on creating storylines with the following features individually: \textit{Luminance, Color histogram, Color moment, Entropy, \#Edges, pHash, Concepts, CNN Dense}, see Figure~\ref{fig:transition-examples2}. 

Figure~\ref{fig:transitions_performance} and Figure~\ref{fig:transition-examples2} show the performance of the proposed transitions baselines on the task of illustrating the EdFest and TDF storylines. The performance was measured by averaging the sum of the scores given by the annotators to each pairwise transition.
All baselines use the same pool of manually selected relevant visual content, for each segment.

In this experiment the \textit{CNN Dense} baseline was one of the three best performing baselines, highlighting the importance of taking into account semantics when optimizing the  quality of transitions. 
The \textit{CNN Dense} baseline, minimises distance between representations extracted from the penultimate layer of the visual concept detector. 
Thus, it is interesting to note that using single concepts as is the case for the \textit{Visual concepts} baseline provides very poor results, stressing the importance of considering other criteria. 

Regarding visual aesthetic features, the best performing baselines where the ones that focus on minimizing the color difference between sequential images in a storyline: \textit{Color histograms} and \textit{Color moment}. 
This supports the assumption that illustrating storylines using content with similar color palettes is a solid way to optimize the quality of visual storylines. 
Now turning to the \textit{Luminance} and \textit{pHash} baselines, these presented varying results, not always being able to ensure high quality transitions. 
Regarding luminance, its important to note that two images can be very distinct while still presenting the same overall luminance value.
Conversely, illustrating storylines by selecting images with similar entropy and number of edges, using the \textit{Entropy} and \textit{\#Edges} baselines, provided worst results. 
This happens because the aesthetic similarities between the sequential images presented in these storylines are, most of the times, not easily perceptible to the naked eye.

\subsection{Learning Multimodal Story Transitions}
\label{sec:Transition_results}
The user study showed that there is not a single feature that can produce the best quality transitions in all situations.
Hence, we trained the Gradient Boosted regressor, section~\ref{sec:transtions} to, given a \textit{transition distance} of a pair of images, predict its quality score, according to the ground truth. 
The \textit{transition distances} for the aforementioned pairs were calculated, using the features defined in Table~\ref{tab:transition-features}. 
The resulting values where then standardized. 
Dividing the resulting dataset into train and test sets, we first trained the Gradient Boosted Trees model with 70\% of the data available. 

\subsection{User Evaluation: Cohesive Storylines}
In this section we evaluated the full framework in a realistic setting with a corpus of ~35k documents and a standard retrieval method (included in our full framework) - hence, there may be non-relevant content in the final visual storyline.
The media editor provides the story segments and the framework computes the optimal illustrations under a transition function trained as described in section~\ref{sec:Transition_results}.
To do so, we considered a completely new set of EdFest and TDF images and stories. The parameters $\beta$ and $\gamma$ were set to $0.9$ and $0.4$ respectively.

To illustrate each segment of each story, the top 10 candidate tweets were selected with the standard text retrieval method BM25. In total, 953 distinct tweets were retrieved, resulting in an average of 10 candidate images to illustrate each segment.
Afterwards the 4 baselines where applied, resulting in a total of 112 storylines.
We proceeded to assess the quality of each visual storyline related to each story topic. 
Hence, the 112 distinct visual storylines were presented to 3 annotators. 
For each visual storyline, the annotators were again asked to rate the relevance of the images to the segment they illustrate and the transition quality between each sequential pair of images with a score of 0 (\textit{"bad"}) or 1 (\textit{"good"}).

\begin{table}
  \centering
  \begin{tabular}{l|ccc|ccc}
    \toprule
    & \multicolumn{3}{c|}{EdFest}    & \multicolumn{3}{c}{TDF}                \\
    Baseline & Rel. & Trans. & Qual. & Rel. & Trans. & Qual.\\
    \midrule
    $MaxTransitions$    & \textbf{0.49}&0.72 & 0.51 & 0.56 & 0.81 & 0.56\\
    $MaxTransitRel$   & 0.48&0.71 & 0.50 & 0.55 & 0.78 & 0.54\\
    $MaxCohesion$      & 0.47&\textbf{0.77} & \textbf{0.52} & \textbf{0.62} & \textbf{0.91} & \textbf{0.64}\\
    $MaxCohesionRel$   & 0.42&0.61 & 0.42 & 0.59 & 0.72 & 0.57\\
      \bottomrule
  \end{tabular}
  \caption{Average performance of the graph based storyline generation methods on the task of illustrating the Edinburgh Festival and Tour de France stories.}
  \label{tab:second_transitions_baselines}
  \vspace{-5mm}
\end{table}

\begin{figure}
    \centering
    %  trim={<left> <lower> <right> <upper>}
    \includegraphics[width=\columnwidth,trim={0 40pt 0 0},clip]{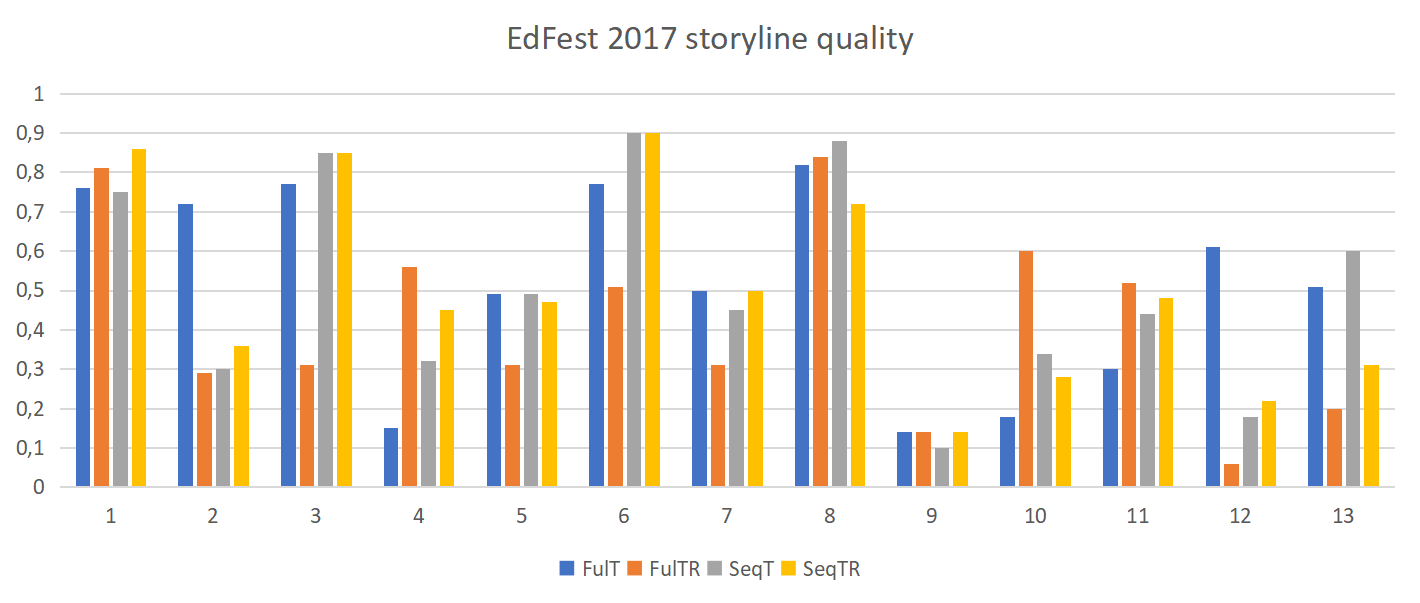}
    \includegraphics[width=\columnwidth,trim={0 40pt 0 0},clip]{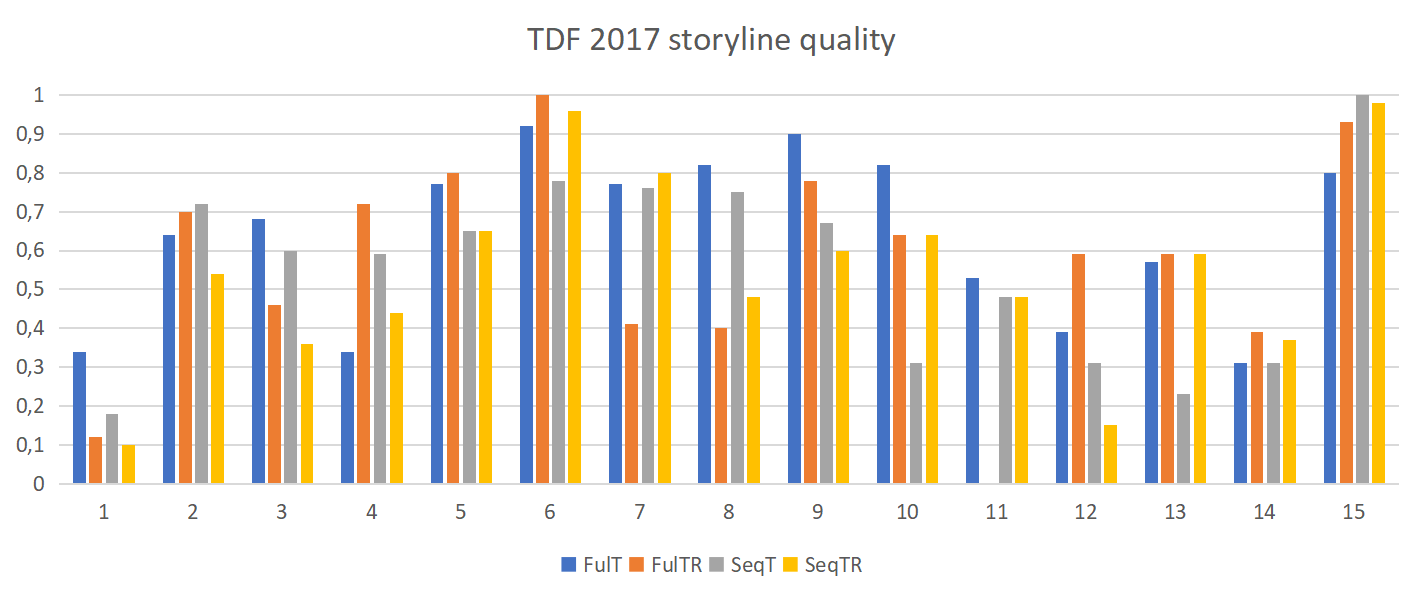}
    \caption{Evaluation of individual storylines for the baselines: 
    \fcolorbox{black}{blue}{\rule{0pt}{4pt}\rule{0pt}{0pt}} MaxCohesion,
    \fcolorbox{black}{BurntOrange}{\rule{0pt}{4pt}\rule{0pt}{0pt}} MaxCohesionRel,
    \fcolorbox{black}{Gray}{\rule{0pt}{4pt}\rule{0pt}{0pt}} MaxTransitions and 
    \fcolorbox{black}{Goldenrod}{\rule{0pt}{4pt}\rule{0pt}{0pt}} MaxTransitionsRel.}
    \label{fig:individual_stories}
    \vspace{-5mm}
\end{figure}

Table~\ref{tab:second_transitions_baselines} shows the performance of the graph based storyline generation methods in terms of the average relevance and transition scores given by the annotators, as well as through the quality assessed by users.
By analyzing them we can verify that, the \textit{maximal-cohesion without relevance} ($MaxCohesion$) approach was the best performing one in terms of average storyline quality, as assessed by the users, but also in terms of transition quality.
Specifically, the storylines created by this approach were rated as having 91\% high quality transitions for the TDF stories, and 77\% high quality transitions for the EdFest stories. This shows that, in a storyline, the transition quality between a pair of images is not just affected by the images of said pair but also by the remaining images in the storyline.
Not only that, but this approach was also the best performing one according to the user assessments, attaining a score of 0.52 for the EdFest stories and a score of 0.64 for the TDF stories.

Furthermore, by inspecting results we verify that the storylines generated by the approaches that leveraged content relevance, $MaxTransitionsRel$ and $MaxCohesionRel$, were not scored as containing more relevant images when compared with the storylines generated by the two remaining approaches. 
By analyzing the storylines individually we see that, the approaches that leverage relevance are, in certain cases, able to pick relevant images to illustrate story segments while the remaining approaches fail to do so. 
However, in turn, seeking higher story cohesion seems to also result in an improved sense of relevance in particular situations.
This happens because, higher transition quality means the images in a storyline tend to present similar semantics. Since stories are related to a particular topic, semantically similar images to the ones already relevant to a story have a high chance of also being relevant to that same story. Consequently, the approaches that leverage content relevance were not able to outperform the remaining ones in terms of finding relevant content.

Figure~\ref{fig:individual_stories} presents the performance of the graph approaches at illustrating individual stories. By analyzing these results we can see that the performance is sensitive to the story being illustrated. Although the best average performing approach was $MaxCohesion$, in a significant number of cases, the storylines created by the other approaches were scored higher. This proves the importance of having different methods of storyline creation, as they provide different illustration alternatives that news editors can work with and build upon. Taking this into account, we can state that in 10 out of 15 TDF stories, at least one of the approaches was able to provide a storyline that was rated with a score near or above 0.7, according to the quality metric. Additionally, in 8 out of 13 EdFest stories, the same can be stated for a score near or above 0.6.

\begin{figure}
    \centering
    \includegraphics[width=\linewidth]{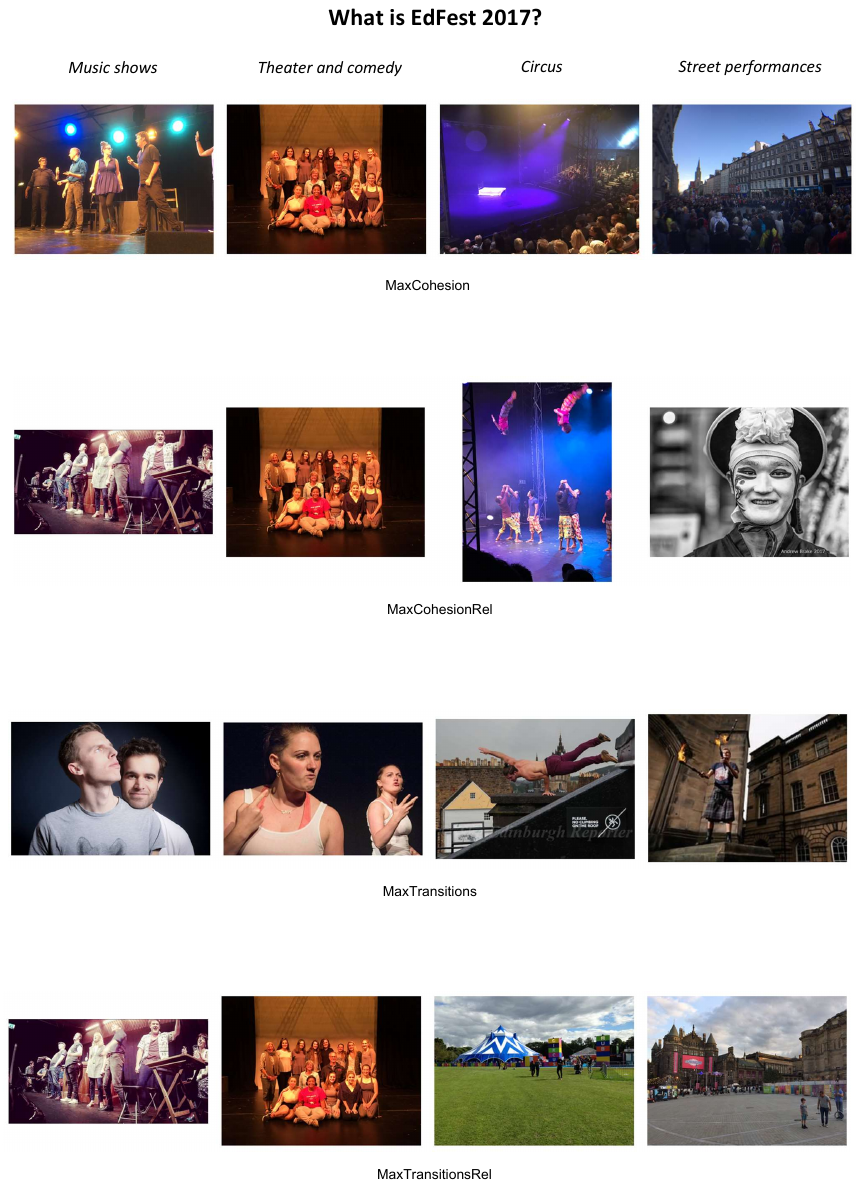}
    \caption{{Illustrations of the ``What is Edinburgh Festival''. It is noticeable that shortest path solutions maximize local similarities (the bottom two stories have very strong relations for image transitions (1, 2) and (3, 4), and low relation for the image pair (2, 3). In contrast, in the top two story illustrations, all images are more homogenous and centered in its semantics (people are always present).}
    }
    \vspace{-5mm}
    \label{fig:final_stories.pdf}
\end{figure}

Observing the examples in Figure~\ref{fig:final_stories.pdf}, one can get a deep insight of how the different storylines created by each approach illustrated the ``\textit{What is Edinburgh Festival}'' story. 
This compares favourably to Figure~\ref{fig:transition-examples2} examples of stories illustrated by the \textit{Color histogram}, \textit{Entropy}, \textit{CCN Dense} and \textit{\#Edges} baselines.
More storylines created by the graph approaches to illustrate the EdFest and TDF stories are available for inspection.
\footnote{ \url{https://run.unl.pt/handle/10362/66267}}.

\section{Conclusion}
In this paper we proposed and formalized a novel computational approach to assist media editors in producing visually cohesive news storylines. Through it, it is possible to express the visual and semantic relationships present in said pairs, in a non-subjective manner. Consequently, this is an important step towards the research of visual news storyline co-creation from a computational perspective. In this context, the key contributions are:
\begin{itemize}
    \item \textbf{Story graphs}: We propose four distinct graph based approaches to visual storyline creation, Figure~\ref{fig:graph}. These approaches proved to be successful at creating high quality visual news storylines, while also providing a solid baseline for future research related to this novel task.
    
    \item \textbf{Learning story transitions:} Leveraging this novel definition, we study the impact of semantic and visual characteristics in transition quality and propose and test a method for predicting the quality of transitions using a Gradient Boosted Trees regressor.
    This approach presented a good performance, proving that, although transition quality is a subjective topic, it is possible to systematically and accurately predict transition quality in an automated manner.
    This could only be achieved by providing the model with the large set of the carefully picked, low and high level features.

    \item \textbf{Cohesive storyline illustration}: The proposed methods provided novel insights into what impacts the perception of news storyline illustrations. We conclude that: i) the \textbf{best perceived quality} of a storyline illustration is achieved when targeting maximum cohesion among all story illustrations; and ii) maximum cohesion also leads to a \textbf{better sense of relevance} of the news story illustration.
\end{itemize}

Finally, although the four approaches to storyline generation presented different performances during evaluation, in the context of news illustration they can all be considered together. They present four alternative illustrated storylines, that can later be reviewed and polished by a news media editor.

\vspace{3mm}
\noindent
\textbf{Acknowledgements.} This work has been partially funded by the CMU Portugal research project GoLocal Ref. CMUP-ERI/TIC/0033/ 2014, by the H2020 ICT project COGNITUS with the grant agreement No 687605 and by the project NOVA LINCS Ref. UIDP/04516/2020. We also gratefully acknowledge the support of NVIDIA Corporation with the donation of the GPUs used for this research.

\balance
\bibliographystyle{ACM-Reference-Format}
\bibliography{library}  

\end{document}